\documentclass[10pt]{article}
\usepackage{CJK}
\usepackage[dvips]{graphicx}
\usepackage{accents}
\usepackage{amssymb}
\usepackage{array}
\usepackage[numbers,sort&compress]{natbib}
\begin{document}

\begin{center}
\textbf{{\large Agegraphic dark energy with the sign-changeable interaction in non-flat universe
}}
 \vskip 0.35 in
\begin{minipage}{4.5 in}
\begin{center}
{\small Y. D. XU
\vskip 0.06 in \textit{ School~of~Science,
\\~Huaihai~Institute~of~Technology,~Lianyungang~222005,~China
\\
ydxu@hhit.edu.cn}}
\end{center}
\vskip 0.2 in

{\small In this paper, we investigate the agegraphic dark energy (ADE) model by including the sign-changeable interaction between ADE and dark matter in non-flat universe. It is shown that the interaction induces an energy flow of which the direction is first from dark matter to ADE and then from ADE to dark matter. The phase space analysis is made and the critical points are found, one of which is the accelerated scaling attractor solution. So, the coincidence problem can be alleviated. Furthermore, we show the evolution of the density parameter $\Omega$, the deceleration parameter $q$ and the equation of state (EoS) parameter $w_{d}$ of ADE. We also find that our model is consistent with the observational data.

\vskip 0.2 in
\textit{Keywords:}  Dark energy; Agegraphic; Sign-changeable interaction.
\\
\\
PACS numbers: 95.36.+x, 98.80.Qc, 98.80.-k}
\end{minipage}
\end{center}

\vskip 0.2 in
\begin{flushleft}\textbf{1. Introduction}\end{flushleft}

Nowadays it is believed that our universe is accelerating, which is supported by cosmological observations \cite{r1,r2,r3,r4}. This cosmic acceleration is commonly explained by dark energy with negative pressure. The simplest dark energy model is Einstein's cosmological constant model ($\Lambda$CDM) with EoS $w_{\Lambda}=-1$. Though the $\Lambda$CDM can fit the observational data well, the scenario suffers from two serious issues called "coincidence problem" and "fine-tuning problem". Among different candidates for dark energy, the ADE model proposed by Cai \cite{r5} has arisen people's great interest recently. This model is based on the uncertainty relation of quantum mechanics as well as the gravitational effect in general relativity. From quantum fluctuations of spacetime, the time parameter $t$ in Minkowski spacetime is not more accurate than $\delta t=\eta t^{2/3}_{p}t^{1/3}$ (Karolyhazy relation) \cite{r6} where $\eta$ is a dimensionless constant of order unity. Based on Karolyhazy relation, the energy density of metric fluctuations of the Minkowski spacetime is given by \cite{r7,r8}
\begin{equation}\label{density}
\rho_{d}\sim\frac{1}{t^{2}_{p}t^{2}}\sim\frac{m^{2}_{p}}{t^{2}},
\end{equation}
where $t_{p}$ is the reduced Planck time. On these basis, the energy density of ADE is given by \cite{r5}
\begin{equation}\label{agegraphic density}
\rho_{d}=\frac{3n^{2}m_{p}^{2}}{T^{2}}.
\end{equation}
Here $m_{p}=(8\pi G)^{-1/2}$, $n$ is a constant parameter and $T$ is chosen to be the age of the universe
\begin{equation}\label{age}
T=\int^{a}_{0}\frac{da}{Ha},
\end{equation}
where
 $a$ is the scale factor, $H\equiv\dot{a}/a$ is the Hubble parameter and a dot denotes the derivative with respect to the cosmic time. Furthermore, the interacting ADE has been proposed and investigated \cite{r9}. It was shown that the EoS of interacting ADE can cross the phantom divide $w_{d}=-1$ from bottom to top. The interacting ADE model also has been extended to the universe with spatial curvature \cite{r10}.

Since the nature of dark energy and dark matter remains unknown, it will not be possible to derive the precise form of the
interaction from fundamental theory. One has to discuss it to a
phenomenological level. The most familiar form of interaction \cite{r9,r10,r11,r12,r13,r14,r15,r16,r17,r18,r19,r20,r21,r22} between dark energy and dark matter is $Q=3cH\rho$. Here $c$ is a coupling constant and positive $c$ means that dark energy decays into dark matter, while negative $c$ means dark matter decays into dark energy, and $\rho$ is taken to be the density of dark matter, dark energy, or the sum of them. Obviously, these interactions are either positive or negative and hence can not change their signs. However, in Ref.\cite{r23}, in a way independent of specific phenomenological form of interaction the authors fitted the interaction term $Q$ with observations. It was found that $Q$
was likely to cross the noninteracting line ($Q = 0$), namely the sign of interaction $Q$
changed in the redshift range of $0.45\leq z\leq0.9$. It is interesting and enlightening to consider
the possibility that the interaction
between dark energy and dark matter changes sign during the cosmological
evolution. Based on this, Wei \cite{r24,r25} proposed a sign-changeable interaction as follow
\begin{equation}\label{sign-changeable interaction}
Q=q(\alpha\dot{\rho}+3\beta H\rho),
\end{equation}
where $\alpha$ and $\beta$ are both dimensionless constants. $q\equiv-\ddot{a}a/\dot{a}^{2}$ is the deceleration parameter.
It is easy to find that the interaction $Q$ can change its sign when the expansion of our universe
changes from deceleration $(q>0)$ to acceleration $(q<0)$. Wei \cite{r24} considered the cosmological evolution of quintessence and phantom with this type of interaction and found this type of interaction can bring new features to cosmology. In Refs. \cite{r26,r27,r28}, it was also shown that the interaction between dark energy and dark matter can change sign during the cosmological evolution.

On the other hand, it is well known that the flatness of the universe is one of the important predictions of conventional inflationary cosmology. The inflation models theoretically produce $\Omega_{\kappa0}$ on the order of $10^{-5}$. But recently Ade et al. \cite{r29} (Planck data) found following constraints on $\Omega_{\kappa0}$
 \begin{eqnarray*}
 100\Omega_{\kappa0}&=&-4.2^{+4.3}_{-4.8}   (\textrm{Planck+WP+highL}),\\
 100\Omega_{\kappa0}&=&-1.0^{+1.8}_{-1.9}  (\textrm{Planck+lensing+WP+highL}).
 \end{eqnarray*}
 These constraints are improved substantially by the addition of BAO data which are
  \begin{eqnarray*}
 100\Omega_{\kappa0}&=&-0.05^{+0.65}_{-0.66}   (\textrm{Planck+WP+highL+BAO}),\\
 100\Omega_{\kappa0}&=&-0.10^{+0.62}_{-0.65}  (\textrm{Planck+lensing+WP+highL+BAO}).
 \end{eqnarray*}
In addition, since the spatial curvature is degenerate with the parameters of dark energy, it is of great importance to study dark energy models with spatial curvature.

The above discussion motivates us to study ADE model by including the sign-changeable interaction (4) between ADE and dark matter in a non-flat universe. The paper is organized as follows: In Sect. 2, we make the phase space analysis and discuss the accelerated scaling attractor solution in the interacting ADE model. In Sect. 3, we show consistency of the model with the observational data. The conclusions are given in Sect. 4.

\begin{flushleft}\textbf{2. Dynamical analysis of the interacting ADE model in a non-flat universe}\end{flushleft}

We consider the Friedmann-Robertson-Walker (FRW) metric for the non-flat universe as
\begin{equation}\label{metric}
ds^{2}=dt^{2}-a^{2}(t)\Big(\frac{dr^{2}}{1-\kappa r^{2}}+r^{2}d\Omega^{2}\Big),
\end{equation}
\noindent where $\kappa=-1,0,1$ is curvature parameter corresponding to open, flat and closed universe, respectively. In our scenario, the non-flat FRW universe contains two components, one is ADE component $\rho_{d}$ and the other is dark matter component $\rho_{m}$ with $w_m=0$, ie., the total energy density $\rho_{tot}=\rho_{d}+\rho_{m}$. The corresponding Friedmann equation is given by
\begin{equation}\label{friedmann}
H^{2}+\frac{\kappa}{a^{2}}=\frac{1}{3m_{p}^{2}}(\rho_{m}+\rho_{d}).
\end{equation}

We assume that the ADE and dark matter exchange energy through
an interaction term $Q$, namely
\begin{eqnarray}\label{continuity}
\dot{\rho}_d+3H(1+w_d)\rho_d&=&-Q,\\
\dot{\rho}_m+3H\rho_m&=&Q,
\end{eqnarray}
where $w_{d}$ is the EoS parameter of ADE. Since the term
$\alpha\dot{\rho}$ in (4) is introduced from the dimensional point of view
\cite{r24}, one can remove this term by setting $\alpha = 0$, and
then (4) becomes simply $Q=3\beta qH\rho$. In this paper, we take the form
\begin{equation}\label{ new interact}
Q=3\beta qH\rho_{d}
\end{equation}
with $\beta$ being a coupling constant. $q$ is the deceleration parameter
\begin{equation}\label{deceleration}
q\equiv-\frac{\ddot{a}a}{\dot{a}^{2}}=-1-\frac{\dot{H}}{H^{2}}.
\end{equation}
Obviously, the interaction term can change its sign when the expansion of our universe changes from deceleration $(q>0)$ to acceleration $(q<0)$.

In what follows, we investigate the dynamical system by phase space analysis and numerically calculate the cosmological evolution. For the phase space analysis in ADE, see \cite{r30,r31}. We introduce the following dimensionless variables
\begin{equation}\label{varialbe}
x=\sqrt{\frac{\rho_{d}}{3m_{p}^{2}H^{2}}},~~~~y=\sqrt{\frac{\rho_{m}}{3m_{p}^{2}H^{2}}},~~~~ \lambda=\frac{\kappa}{H^{2}a^{2}}.
\end{equation}
Differentiating Eq.(6) with respect to the cosmic time and combing Eqs.(2), (8), (9) and (10), one can easily get
\begin{equation}\label{ differ H}
\frac{\dot{H}}{H^{2}}=\frac{\lambda-\frac{3}{2}y^{2}-x^{2}(\frac{x}{n}+\frac{3}{2}\beta)}{1+\frac{3}{2}\beta x^{2}}.
\end{equation}
From Eqs.(11) and (12), we have the following autonomous system
\begin{eqnarray}\label{auto}
x'&=&x(q+1-\frac{x}{n}),\\
y'&=&-\frac{y}{2}+(y+\frac{3}{2}\beta\frac{x^{2}}{y})q,\\
\lambda'&=&2q\lambda,
\end{eqnarray}
where
\begin{equation}\label{dec}
q=-1-\frac{\dot{H}}{H^{2}}=\frac{-1-\lambda+\frac{3}{2}y^{2}+\frac{1}{n}x^{3}}{1+\frac{3}{2}\beta x^{2}}
\end{equation}
and the prime denote the derivative with respect to $N=\ln a$. From Eq.(6), we have
\begin{equation}\label{friedmann2}
x^{2}+y^{2}=1+\lambda.
\end{equation}
The fractional energy density of the ADE, dark matter and curvature is
\begin{eqnarray}\label{fractional}
\Omega_{d}&\equiv&\frac{\rho_{d}}{3m_{p}^{2}H^{2}}=x^{2},~~~~~\Omega_{m}\equiv\frac{\rho_{m}}{3m_{p}^{2}H^{2}}=y^{2}, \nonumber \\
\Omega_{\kappa}&\equiv&\frac{\kappa}{H^{2}a^{2}}=\lambda.
\end{eqnarray}
The EoS of ADE could be expressed in terms of the new variables as
\begin{equation}\label{eos}
w_{d}=-1+\frac{2x}{3n}-\frac{\beta(-1-\lambda+\frac{3}{2}y^{2}+\frac{1}{n}x^{3})}{1+\frac{3}{2}\beta x^{2}}.
\end{equation}
In the absence of interaction ($\beta=0$), from Eq.(19), one can see that $w_{d}$ is always larger than -1 and cannot cross the phantom divide $w_{d}=-1$. However, the situation is changed when the interaction term is
taken into account. In this case $(\beta\neq0)$, from Eq.(19), it is easy to see that $w_{d}$ can cross the phantom divide provided
\begin{equation}\label{condition}
\frac{2x}{3n}<\frac{\beta(-1-\lambda+\frac{3}{2}y^{2}+\frac{1}{n}x^{3})}{1+\frac{3}{2}\beta x^{2}}.
\end{equation}

Setting $x'=y'=\lambda'=0$ in Eqs. (13), (14) and (15), we can obtain the physically meaningful critical points $(x_{c},y_{c},\lambda_{c})$ of the autonomous system. The critical points and their properties are given in the Table 1. The first critical point, $(0,1,0)$ is unstable and corresponds to a matter dominated phase. The other critical point, $(x_{*},y_{*},0)$ is the attractor. $x_{*}$ satisfies
\begin{equation}\label{condition1}
\frac{3\beta-2}{n}x^{3}+(3-3\beta)x^{2}+\frac{2x}{n}-3=0
\end{equation}
and $y_{*}=\sqrt{1-x^{2}_{*}}$. The deceleration parameter and the EoS of ADE in the attractor are equal respectively $q_{*}=-1+\frac{x_{*}}{n}$ and $w_{d*}=-1+\beta+(\frac{2}{3}-\beta)\frac{x_{*}}{n}$. For example, consider the case when $n=3.5$, $\beta=-0.3$. We can find that in this model $q_{*}\approx-0.75$ and $w_{d*}\approx-1.05$.
In the absence of interaction ($\beta=0$), from Eq.(21), one can get $x_{*}=1$, which corresponds to ADE dominated phase. Eq.(21) can be also expressed as
\begin{equation}\label{condition2}
\big[(3\beta-2)x^{2}+2\big](\frac{x}{n}-1)=1-x^{2}.
\end{equation}
Since $x_{*}^{2}\leq1$, from Eq.(22) we can find $\beta\leq0$ for $n>1$. It means that the interaction term (9) can change its sign from $Q<0$ to $Q>0$ when the expansion of our universe changes from deceleration $(q>0)$ to acceleration $(q<0)$.
If the attractor $(x_{*},y_{*},0)$ is the scaling solution, the coincidence problem is alleviated. Scaling solutions are characterized by a constant dark matter to dark energy ration $r=y^{2}/x^{2}$. We can find that
\begin{equation}\label{ratio}
r'=3\beta q-(3-\frac{2x}{n})r.
\end{equation}
The scaling solutions mean $r'=0$, which results in $r_{*}=3\beta(x_{*}-n)/(3n-2x_{*})$. So the attractor $(x_{*},y_{*},0)$ is just the scaling solution for $(\beta<0)$. Thus the coincidence problem can be alleviated because, regardless of the initial conditions, the system evolves toward a final state where the ratio of dark matter to dark energy stays constant. In Fig.1, it is easy to find that $r_{*}\approx0.27$. In addition, since $x_{*}^{2}\leq1$, we can see $q_{*}<0$ for $n>1$. So the critical point $(x_{*},y_{*},0)$ is the accelerated scaling attractor solutions.

In Fig.2, we show the evolution of $\Omega_{d}$ and $\Omega_{m}$ with respect to $N=\ln a$. We fix $n=0.35$ and let $\beta$ vary. From Fig.2, one can see that increasing the value of $|\beta|$ results in lager $\Omega_{m}$ and smaller $\Omega_{d}$ in the late time.

Fig.3 shows the evolution of the deceleration parameter $q$ with respect to $N=\ln a$ for different $\beta$ and the fixed $n$. From Fig.3, we can see that the universe has a transition from deceleration $(q>0)$ to acceleration $(q<0)$. In addition, the model parameter may affect the time of onset of the acceleration. For the larger $|\beta|$ the acceleration sets in earlier.

In Fig.4, the evolution of EoS of ADE $w_{d}$ is shown for different $\beta$ and the fixed $n$. We find that $w_{d}$ is more likely to cross the phantom divide $w_{d}=-1$ from top to bottom with the increasing of the $|\beta|$ in our model. Whereas in Ref.\cite{r9}, the EoS of interacting ADE can cross the phantom divide from bottom to top.

\begin{table}
\caption{Location of the critical points of the autonomous system of Eqs. (13-15), their stability and dynamical behavior of the Universe at those points. }
\begin{center}
\begin{tabular}{ c  c c c  }
  \hline

  $(x_{c},y_{c},\lambda_{c})$  & Stability  & $q$ & $w_{d}$  \\
  coordinates & character & & \\
  \hline
  (0,1,0) & unstable & 1/2 & $-1-\beta/2$ \\
  $(x_{*},y_{*},0)$ & attractor & $q_{*}<0$ & $w_{d*}$ \\
  \hline
\end{tabular}
\end{center}
\end{table}

\begin{figure}
\begin{center}
  \includegraphics[width=0.4\textwidth]{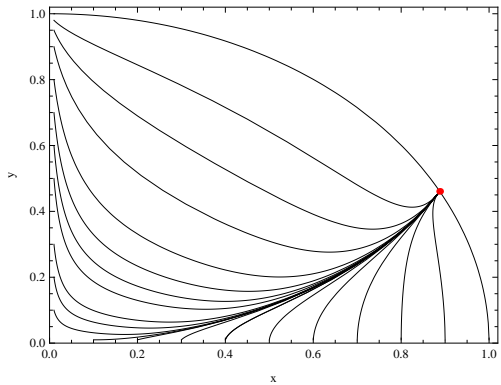}

\caption{The phase plane for $n=3.5$ and $\beta=-0.3$. The red dot stands for the late-time attractor with $x_{*}\approx0.89$, $y_{*}\approx0.46$ ($\Omega_{d*}\approx0.79$, $\Omega_{m*}\approx0.21$).}
\label{fig:1}
\end{center}
\end{figure}

\begin{figure}
\begin{center}
  \includegraphics[width=0.4\textwidth]{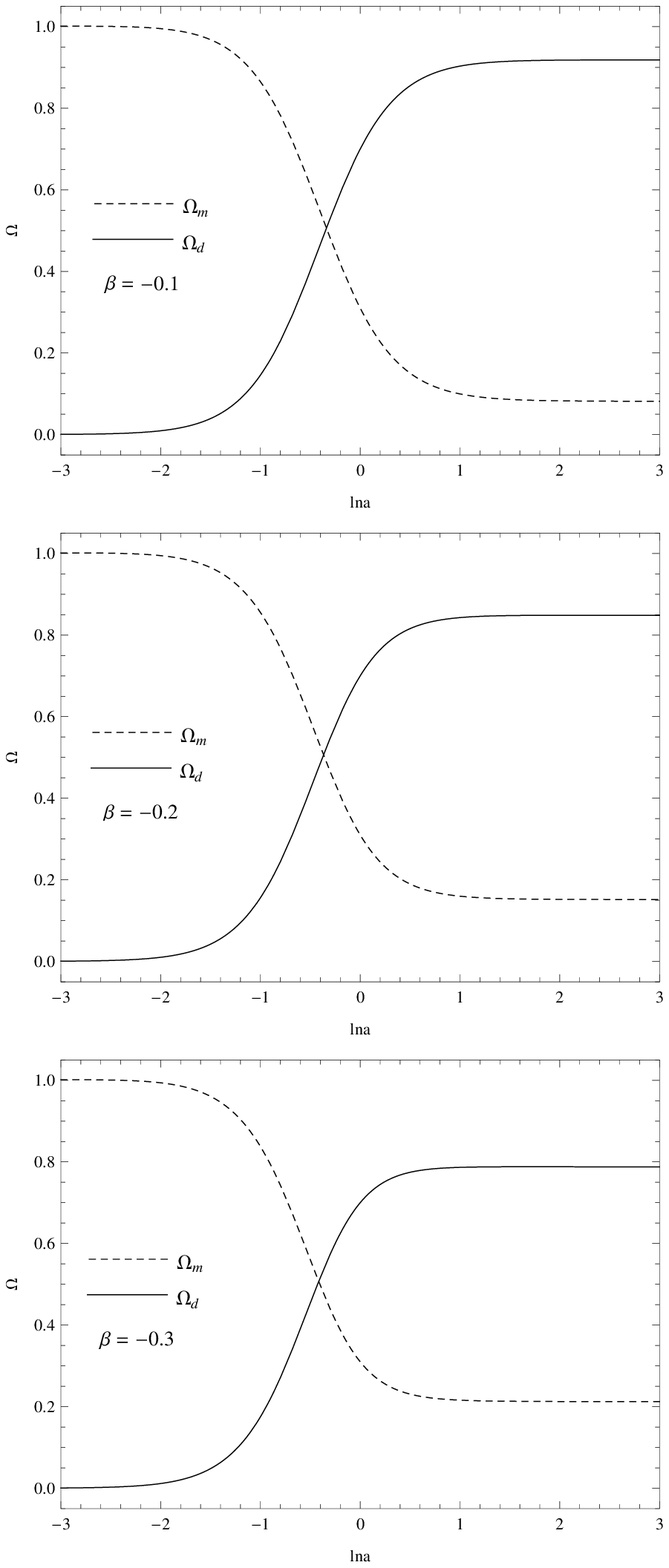}

\caption{The evolution of $\Omega_{d}$ (solid line) and $\Omega_{m}$ (dot line) for $n=3.5$ and $\beta$ as indicated.}
\label{fig:2}
\end{center}
\end{figure}

\begin{figure}
\begin{center}
  \includegraphics[width=0.4\textwidth]{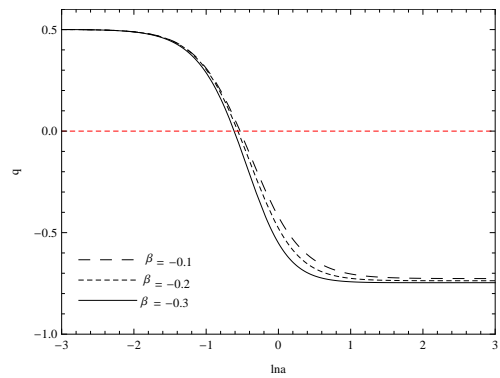}

\caption{The evolution of deceleration parameter $q$ for $n=3.5$ and $\beta$ as indicated.}
\label{fig:3}
\end{center}
\end{figure}

\begin{figure}
\begin{center}
  \includegraphics[width=0.4\textwidth]{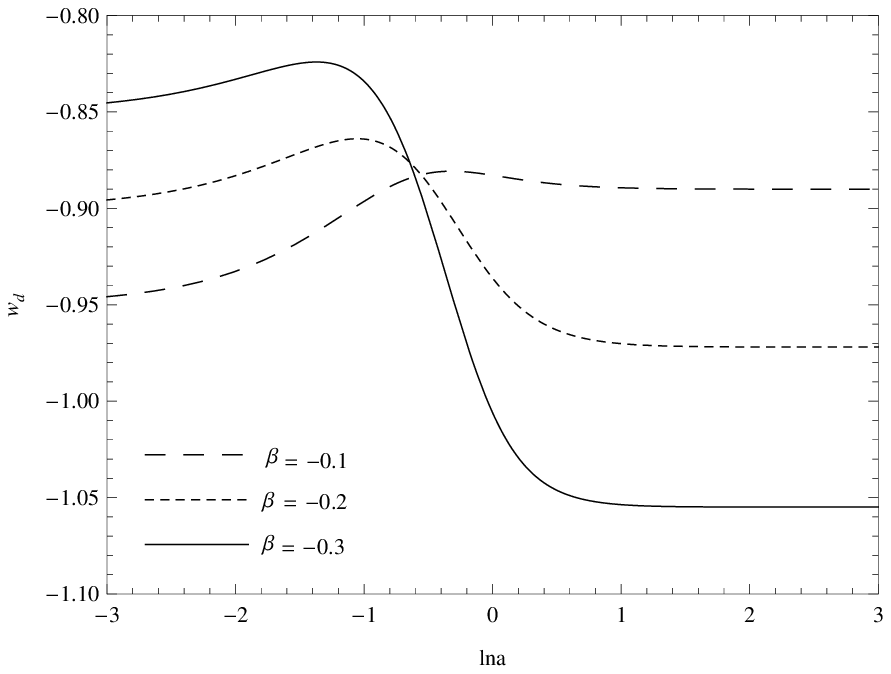}

\caption{The evolution of the EoS of ADE $w_{d}$ for $n=3.5$ and $\beta$ as indicated.}
\label{fig:4}
\end{center}
\end{figure}

\begin{flushleft}\textbf{3. Observational data}\end{flushleft}

Since there is mo fundamental theory which can be used to select a specific interacting dark energy model, any interacting dark energy model will necessarily be phenomenological. There are two criterions to determine whether the model is correct and feasible. One is to examine whether the interacting model can lead to the accelerated scaling attractor solutions, which alleviate the coincidence problem, the other is the observations.  In Sect. 2, it has been shown that the accelerated scaling attractor solutions do exist in our model. In the present section, we will consider the Union2.1 Type Ia supernova dataset (SNeIa) \cite{r32} , which contains 580 data points. We will check the interacting ADE model via the observational data.

The 580 data points of SNeIa are given in terms of the distance modulus $\mu_{obs}(z_{i})$. On the other hand, the theoretical distance modulus is defined as
\begin{equation}\label{distance modulus}
\mu_{th}(z_{i})=5\log_{10}D_{L}(z_{i})+\mu_{0},
\end{equation}
where the zero offset $\mu_{0}$ depends on $H_{0}$ (or $h$) as
\begin{equation}\label{zero}
\mu_{0}=5\log_{10}\big(\frac{cH^{-1}_{0}}{\mathrm{Mpc}}\big)+25=-5\log_{10}h+42.38
\end{equation}
and $h$ is the Hubble constant $H_{0}$ in units of 100 $\mathrm{km~s^{-1} Mpc^{-1}}$. The theoretically predicted value $D_{L}(z)$ in the context of a given model $H(z;\mathbf{p})$ can be described by
\begin{equation}\label{theoretically predicted value}
D_{L}(z)=\frac{1+z}{\sqrt{|\Omega_{\kappa0}|}}\mathrm{Sinn}\big[\sqrt{|\Omega_{\kappa0}|}\int^{z}_{0}dz'\frac{H_{0}}{H(z';\mathbf{p})}\big],
\end{equation}
where $\mathbf{p}$ denotes the model parameters, $\Omega_{\kappa0}=\kappa/(H^{2}_{0}a^{2}_{0})$ and $\mathrm{Sinn}(x)=\sin(x),x,\sinh(x)$ for respectively a spatially closed $\Omega_{\kappa}>0$, flat $\Omega_{\kappa}=0$ and open $\Omega_{\kappa}<0$ universe.

Different models result in different theoretical distance modulus $\mu_{th}$. We can judge the plausibility of an cosmological model by comparing $\mu_{th}$ and the observational value of $\mu_{obs}$. In order to see whether the theoretical model corresponds to the observational data, we must have the value of $H$, which can be obtained through the autonomous system (13-15). From Fig.5, one can see that our model is consistent with the observational data.

\begin{figure}
\begin{center}
  \includegraphics[width=0.4\textwidth]{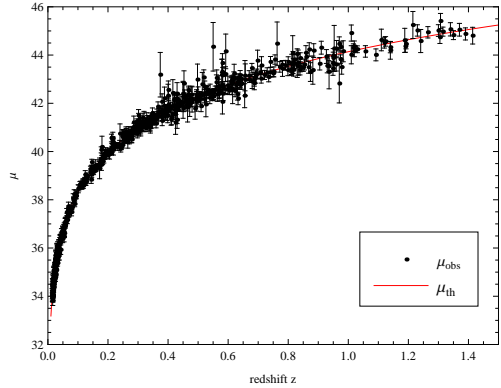}

\caption{The observed 580 SNeIa distance modulus along with the theoretically predicted curves (red solid line) in the model with $n=3.5$ and $\beta=-0.3$, where we take a \textit{priori} that current dimensionless Hubble parameter $h=0.70$.}
\label{fig:5}
\end{center}
\end{figure}

\begin{flushleft}\textbf{4. Conclusions}\end{flushleft}

In the present paper, we investigate ADE model by including the sign-changeable interaction ($Q=3\beta qH\rho_{d}$) between ADE and dark matter in non-flat universe. Using the phase-plane analysis, the dynamical behavior of the model has been studied. It was found that the accelerated scaling attractor solution did exist in the model, which can help to alleviate the coincidence problem. Moreover, we find that the the condition $\beta<0$ is necessary for the existence of the scaling solution. It means that the interaction $Q$ can change its sign from $Q<0$ to $Q>0$ when the expansion of our universe changes from deceleration $(q>0)$ to acceleration $(q<0)$, which is opposite to the case in Ref.\cite{r28}. This indicates that at first dark matter decays to ADE, and then ADE decays to dark matter.  We also show numerically that the coupling constant $\beta$ plays an important role in the evolution of the universe. It is easy to see from Figs.2-4 that for the fixed $n$, increasing the value of $|\beta|$ can bring the features as follows: larger $\Omega_{m}$ is obtained in the late time, the acceleration sets in earlier and  $w_{d}$ is more likely to cross the phantom divide from top to bottom.

Next, using 580 SNeIa data, we checked the interacting ADE model. The model has given a series of reasonable pictures of the cosmic evolution and it is consistent with the observational data. Our work implies that we should pay more attention to a sign-changeable interaction between dark sectors.

\begin{flushleft}{\noindent\bf References}

\bibitem{r1}
S. Perlmutter et al., Astrophys. J. \textbf{517}, 565 (1999)

\bibitem{r2}
C.L. Bennett et al., Astrophys. J. Suppl. \textbf{148}, 1 (2003)

\bibitem{r3}
M. Tegmark et al., Phys. Rev. D \textbf{69}, 103501 (2004)

\bibitem{r4}
S.W. Allen, R.W. Schmidt, H. Ebeling, A.C. Fabian, L. vani Speybroeck, Mon. Not. R. Astron. Soc. \textbf{353}, 457 (2004)

\bibitem{r5}
R.G. Cai, Phys. Lett. B \textbf{657}, 228 (2007)

\bibitem{r6}
F. Karolyhazy, Nuovo.Cim. A \textbf{42}, 390 (1966)

\bibitem{r7}
M. Maziashvili, Int. J. Mod. Phys. D \textbf{16}, 1531 (2007)

\bibitem{r8}
M. Maziashvili, Phys. Lett. B \textbf{652}, 165 (2007)

\bibitem{r9}
H. Wei, R.G. Cai, Eur. Phys. J. C \textbf{59}, 99 (2009)

\bibitem{r10}
A. Sheykhi, Phys. Lett. B \textbf{680}, 113 (2009)

\bibitem{r11}
L. Amendola, Phys. Rev. D \textbf{60}, 043501 (1999)

\bibitem{r12}
L. Amendola, Phys. Rev. D \textbf{62}, 043511 (2000)

\bibitem{r13}
L. Amendola, C. Quercellini, Phys. Rev. D \textbf{68}, 023514 (2003)

\bibitem{r14}
W. Zimdahl, D. Pavon, L.P. Chimento, Phys. Lett. B \textbf{521}, 133 (2001)

\bibitem{r15}
D. Pavon, B. Wang, arXiv:0712.0565

\bibitem{r16}
L.P. Chimento, A.S. Jakubi, D. Pavon, W. Zimdahl, Phys. Rev. D \textbf{67}, 083513 (2003)

\bibitem{r17}
Z.K. Guo, R.G. Cai, Y.Z. Zhang, JCAP \textbf{0505}, 002 (2005)

\bibitem{r18}
Z.K. Guo, N. Ohta, S. Tsujikawa, Phys. Rev. D \textbf{76}, 023508 (2007)

\bibitem{r19}
R.G. Cai, A. Wang, JCAP \textbf{0503}, 002 (2005)

\bibitem{r20}
B. Wang, J. Zang, C.Y. Lin, E. Abdalla, S. Micheletti, Nucl. Phys. B \textbf{778}, 69 (2007)

\bibitem{r21}
J. H. He, B. Wang, JCAP \textbf{0806}, 010 (2008)

\bibitem{r22}
S.H. Pereira, J.F. Jesus, Phys. Rev. D \textbf{79}, 043517 (2009)

\bibitem{r23}
R.G. Cai, Q. Su, Phys. Rev. D \textbf{81}, 103514 (2010)

\bibitem{r24}
H. Wei, Nucl. Phys. B \textbf{845}, 381 (2011)

\bibitem{r25}
H. Wei, Commun. Theor. Phys. \textbf{56}, 972 (2011)

\bibitem{r26}
C.Y. Sun, R.H. Yue, Phys. Rev. D \textbf{85}, 043010 (2012)

\bibitem{r27}
Y. H. Li, X. Zhang, Eur. Phys. J. C \textbf{71}, 1700 (2011)

\bibitem{r28}
J.F. Zhang, Y.Y. Li, Y. Liu, S. Zou, X. Zhang, Eur. Phys. J. C \textbf{72}, 2077 (2012)

\bibitem{r29}
P.A.R. Ade et al., arXiv:1303.5076

\bibitem{r30}
O.A. Lemets, D.A. Yerokhin, L.G. Zazunov, JCAP \textbf{01}, 007 (2011)

\bibitem{r31}
X.M. Liu, Z.X. Zhai, K. Xiao, W.B. Liu, Eur. Phys. J. C \textbf{72}, 2057 (2012)

\bibitem{r32}
N. Suzuki, D. Rubin, C. Lidman et al., Astrophys. J. \textbf{746}, 85 (2012)

\end{flushleft}

\end{document}